\documentclass[11pt]{article}
\usepackage{moriond,epsfig,psfrag}

\def\be{\begin{equation}}
\def\ee{\end{equation}}
\def\bea{\begin{eqnarray}}
\def\eea{\end{eqnarray}}

\psfrag{log(mN)}{$\log{m_N}$}
\psfrag{BR(tau -> 3mu)}{BR($\tau \to 3 \mu$)}
\psfrag{BR(tau -> mu + gamma)}{BR($\tau \to \mu \gamma$)}
\psfrag{BR(tau -> 3e)}{BR($\tau \to 3 e$)}
\psfrag{BR(tau -> e + gamma)}{BR($\tau \to e \gamma$)}
\psfrag{BR(tau -> e +gamma)}{BR($\tau \to e \gamma$)}
\psfrag{BR(mu -> 3e)}{BR($\mu \to 3 e$)}
\psfrag{BR(mu -> e + gamma)}{BR($\mu \to e \gamma$)}
\psfrag{tanb}{$\tan{\beta}$}
\psfrag{tan b}{$\tan{\beta}$}
\psfrag{mod(theta1)}{$\vert \theta_1 \vert$}
\psfrag{mod(theta2)}{$\vert \theta_2 \vert$}
\psfrag{mod(theta3)}{$\vert \theta_3 \vert$}
\psfrag{mod(theta13)}{$\vert \theta_{13} \vert$}
\psfrag{M0}{$M_0$ (GeV)}
\psfrag{M1/2}{$M_{1/2}$ (GeV)}
\psfrag{M0=M1/2}{$M_0 = M_{1/2}$ (GeV)}
\psfrag{Deltas12}{$|\delta_{12}|$}
\psfrag{Deltas13}{$|\delta_{13}|$}
\psfrag{Deltas23}{$|\delta_{23}|$}
\psfrag{theta13}{$\theta_{13}$}
\psfrag{Ynu21}{$|Y_{\nu}^{12}|$}
\psfrag{Ynu22}{$|Y_{\nu}^{22}|$}
\psfrag{Ynu23}{$|Y_{\nu}^{32}|$}
\psfrag{Ynu31}{$|Y_{\nu}^{13}|$}
\psfrag{Ynu32}{$|Y_{\nu}^{23}|$}
\psfrag{Ynu33}{$|Y_{\nu}^{33}|$}

\begin{document}
\title{Lepton Flavor Violating $\tau$ and $\mu$ decays}

\author{ E. Arganda, M.J. Herrero}

\address{Departamento de F\'{\i}sica Te\'orica, UAM/IFT, 28049, Madrid, Spain}

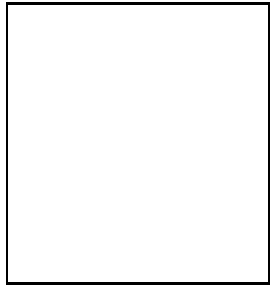
\maketitle\abstracts{
In this work the following lepton flavor violating $\tau$ and $\mu$ decays
are studied: $\tau^- \to \mu^- \mu^- \mu^+$,
$\tau^- \to e^- e^- e^+$, $\mu^- \to e^- e^- e^+$, 
$\tau^- \to \mu^- \gamma$, $\tau^- \to e^- \gamma$ and 
$\mu^- \to e^- \gamma$. We work in a constrained supersymmetric 
scenario with universal soft breaking terms and where the MSSM particle content is extended by the addition of three heavy right handed
Majorana neutrinos and their supersymmetric partners. The 
generation of neutrino masses is done via the seesaw mechanism. We analize all these lepton flavor violating decays 
in terms of the relevant input parameters, which are the usual mSUGRA parameters and the seesaw parameters. In the numerical analysis 
compatibility with the most recent experimental upper bounds on all these 
$\tau$ and $\mu$
decays, with the neutrino data, and with the present lower bounds on the 
supersymmetric particle masses are required. Two typical scenarios with
degenerate and hierarchical heavy neutrinos are considered. 
We will show here that
the minimal supergravity and seesaw 
parameters do get important restrictions from these $\tau$ and $\mu$ 
decays in the
hierarchical neutrino case.}

\section{Introduction}
\label{Intro}

The present strong evidence for lepton flavor changing neutrino 
oscillations  in
neutrino data implies the existence of non-zero masses for the light neutrinos, 
and provides the first experimental clue for physics beyond the 
Standard Model (SM). This fact does require a theoretical framework beyond the SM with just three massless left-handed neutrinos. Within the MSSM-seesaw context, which will be adopted here, the MSSM particle content is enlarged
by three right handed neutrinos plus their corresponding supersymmetric (SUSY) partners, and the 
neutrino masses are generated by the seesaw mechanism. Three of the six resulting Majorana
neutrinos have light masses, $m_{\nu_i}, i=1,2,3$, and the other three have heavy  masses, 
$m_{N_i}, i=1,2,3$. These physical masses can be related to the Dirac mass matrix 
$m_D$, the right-handed neutrino mass matrix $m_M$, and the unitary matrix $U_{MNS}$ by 
$\mbox{diag}(m_{\nu_1}, m_{\nu_2}, m_{\nu_3})\simeq U_{MNS}^T(-m_Dm_{M}^{-1}m_D^T)U_{MNS}$ and 
$\mbox{diag}(m_{N_1}, m_{N_2}, m_{N_3})\simeq m_M$, respectively. Here we have chosen an electroweak 
eigenstate basis where $m_M$ and the charged lepton mass matrix are flavor diagonal, and we
have assumed that all elements in $m_D=Y_{\nu}<H_2>$, where $Y_{\nu}$ is the neutrino Yukawa
coupling matrix and $<H_2>=v \sin \beta$ ($v=174$ GeV), are much smaller than those of $m_M$. The two
previous relations can be rewritten together in a more convenient form for the work presented
here as, 
$m_D^T =i m_N^{\mbox{diag} \, 1/2} R m_{\nu}^{\mbox{diag} \, 1/2} U_{MNS}^+$, where $R$ is a general complex
and orthogonal $3 \times 3$ matrix, which will be parameterized by three complex angles
$\theta_i$, $i=1,2,3$. 

One of the most interesting features of the SUSY-seesaw models is the associated rich
phenomenology due to the occurrence of lepton flavor violating (LFV) processes. In these models the LFV ratios can be large due to an important source of lepton flavor mixing in the soft-SUSY-breaking terms. Even in the
scenarios with universal soft-SUSY-breaking parameters at the large
energy scale associated to the SUSY breaking $M_X$, the running from this scale down to $m_M$ 
induces, via the neutrino Yukawa couplings, large lepton flavor mixing in the slepton 
soft masses, 
and provides the so-called slepton-lepton misalignment, which in turn
generates 
non-diagonal lepton flavor interactions. 
These interactions can induce sizable ratios in several LFV processes which are actually being tested experimentally with high precision and therefore provide 
a very interesting window to look for indirect SUSY signals. Among these 
processes, the LFV $\tau$ and $\mu$ decays are
probably the most interesting ones for various reasons. 
On one hand, they get
vanishing rates
in the SM with massless neutrinos and highly suppressed rates
in the SM with massive netrinos. The smallness of these rates in the
non-SUSY version of the seesaw mechanism for neutrino mass generation 
is due to their suppression by inverse powers of the heavy scale 
$m_M$.      
On the other hand, although these decays have not
been seen so far in the present experiments, there are very 
restrictive upper
bounds~\cite{LFVdata} on their possible rates which imply important restrictions on the 
new physics beyond the SM. These restrictions apply severely to the case
of softly broken SUSY theories with massive neutrinos and the seesaw
mechanism, since these give rise to higher rates, 
being suppressed  
by inverse powers of the SUSY breaking scale, 
$m_{SUSY}\leq 1$ TeV, instead of inverse powers of $m_M$.

Here we compute the partial widths for the the LFV $\tau$ and $\mu$ 
decays of type
$l_j\to l_i \gamma$ and $l_j\to 3l_i$
to one-loop order
and analyze numerically the corresponding branching ratios
in terms of the mSUGRA and seesaw parameters, namely, $M_0$,
$M_{1/2}$, $\tan\beta$, $m_{N_i}$ and $R$. To solve numerically the RGEs we use 
the Fortran code SPheno~\cite{SPheno} that we have adapted to include the full flavor structure of the 
$3 \times 3$ soft SUSY breaking mass and trilinear coupling matrices and of the 
Yukawa coupling matrices. Our final goal is to use the SUSY contributions to LFV $\tau$ and $\mu$ decays 
as an efficient way to test the mSUGRA and seesaw parameters and we explore 
in detail the restrictions imposed from the present experimental bounds. 
We
will find that for some plausible choices of the seesaw parameters,
being compatible with neutrino data, there are indeed large excluded 
regions in the 
mSUGRA parameter space. 

For the numerical analysis, the $U_{MNS}$ matrix elements and the $m_{\nu_i}$ are fixed
to the most favored values by neutrino data with  
$\sqrt{\Delta m_{sol}^2}=0.008$ eV, $\sqrt{\Delta m_{atm}^2}=0.05$ eV, 
$\theta_{12}=\theta_{sol}=30^o$, $\theta_{23}=\theta_{atm}=45^o$, 
$\theta_{13}=0^o$ and $\delta = \alpha= \beta =0$. Some results will also be presented 
for the alternative choice of small but non-vanishing $\theta_{13}$. We consider two plaussible scenarios, 
one with
quasi-degenerate light and degenerate heavy neutrinos and  with    
$m_{\nu_1}=0.2$ eV, $m_{\nu_2}=m_{\nu_1}+\frac{\Delta m_{sol}^2}{2 m_{\nu_1}}$, $m_{\nu_3}=m_{\nu_1}+\frac{\Delta m_{atm}^2}{2 m_{\nu_1}}$ and 
$m_{N_1} = m_{N_2}= m_{N_3}= m_N$; and the other one with hierarchical light and hierarchical
heavy neutrinos, and with 
$m_{\nu_1} \simeq 0$ eV, $m_{\nu_2}= \sqrt{\Delta m_{sol}^2}$, 
$m_{\nu_3}=\sqrt{\Delta m_{atm}^2}$ and
$m_{N_1} \leq  m_{N_2} < m_{N_3}$. 

This is a reduced version of our more complete work~\cite{nosotros} to which we address the
reader for more details.

\section{Numerical results and conclusions}
\label{NumConclu}

For the degenerate heavy neutrinos case, 
we have gotten rates for all the studied LFV $\tau$ and $\mu$ decays that
are below the present experimental upper bounds, and we are basically in agreement with previous results~\cite{Hisano}. 
The largest rates we get, within the explored range of the seesaw and 
mSUGRA parameter space, are for the $\tau$ decays. Specifically, 
BR$(\tau \to \mu \gamma) \sim 10^{-8}$ and 
BR$(\tau^- \to \mu^- \mu^- \mu^+) \sim 3 \times 10^{-11}$, 
corresponding to the extreme values of $\tan{\beta} = 50$ and 
$m_{N} = 10^{14}$ GeV and for the lowest values of $M_0$ and $M_{1/2}$ 
explored. The case of hierarchical heavy neutrinos turns out to be much more interesting.
 First of all, we get much larger branching ratios than in the previous case and
 secondly they are in many cases above the present experimental bounds. 

\begin{figure}
\begin{center}
\psfig{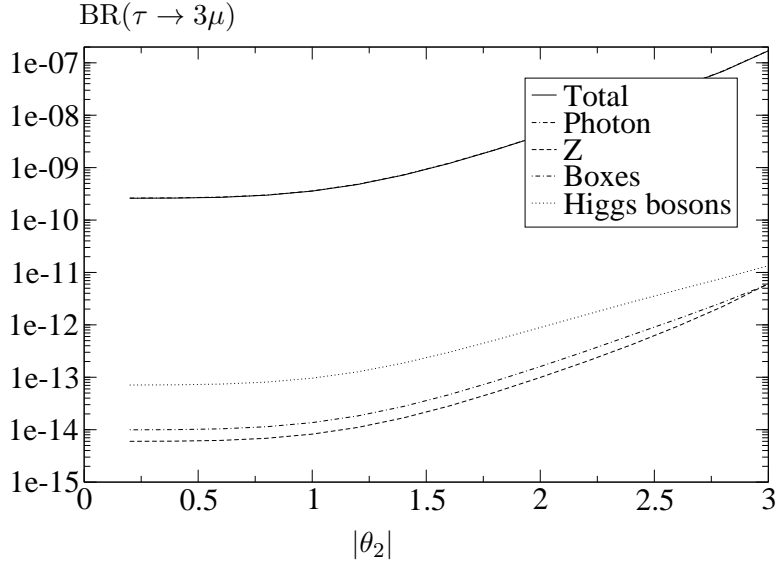}
\end{center}
\caption{Dependence of BR$(\tau^- \to \mu^- \mu^- \mu^+)$ with  
$\vert \theta_2 \vert$ for $\arg(\theta_2) = \pi/4$, $(m_{N_1},m_{N_2},m_{N_3})=(10^8, 
2 \times 10^8, 10^{14})$ GeV, $\theta_1=\theta_3=0$, $\tan \beta = 50$, 
$M_0=400 $ GeV, $M_{1/2}=300 $ GeV, $A_0 = 0$ and $\mbox{sign}(\mu) > 0$.
\label{fig:1}}
\end{figure}

\begin{figure}
\begin{center}
\psfig{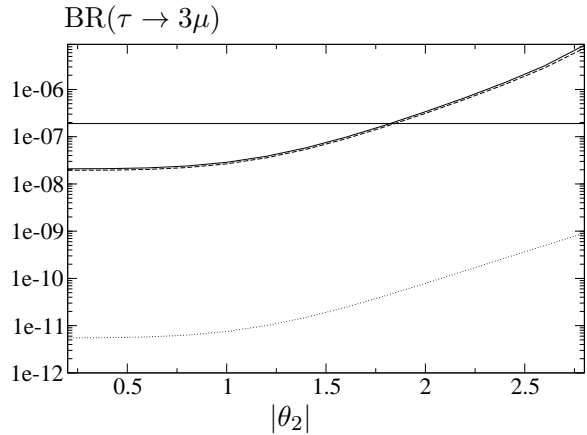}
\end{center}
\caption{Dependence of LFV $\tau \to 3 \mu$  with $|\theta_{2}|$ in
  with hierarchical heavy neutrinos, for different
  $m_{N_i}$ choices. Solid line is for
  $(m_{N_1},m_{N_2},m_{N_3})=(10^8, 2 \times 10^8, 10^{14})$ GeV,
  dashed line is for $(m_{N_1},m_{N_2},m_{N_3})=(10^{10}, 2 \times
  10^{10}, 10^{14})$ GeV, and dotted line is for
  $(m_{N_1},m_{N_2},m_{N_3})=(10^8, 2 \times 10^8, 10^{12})$ GeV. The
  rest of parameters are set to $\tan \beta = 50$, $M_0=200 $ GeV,
  $M_{1/2}=100 $ GeV, $A_0 = 0$, $\mbox{sign}(\mu) > 0$ and
  $\mbox{arg}(\theta_2) = \pi/4$. The horizontal line is the experimental bound.
\label{fig:4}}
\end{figure}

\begin{figure}
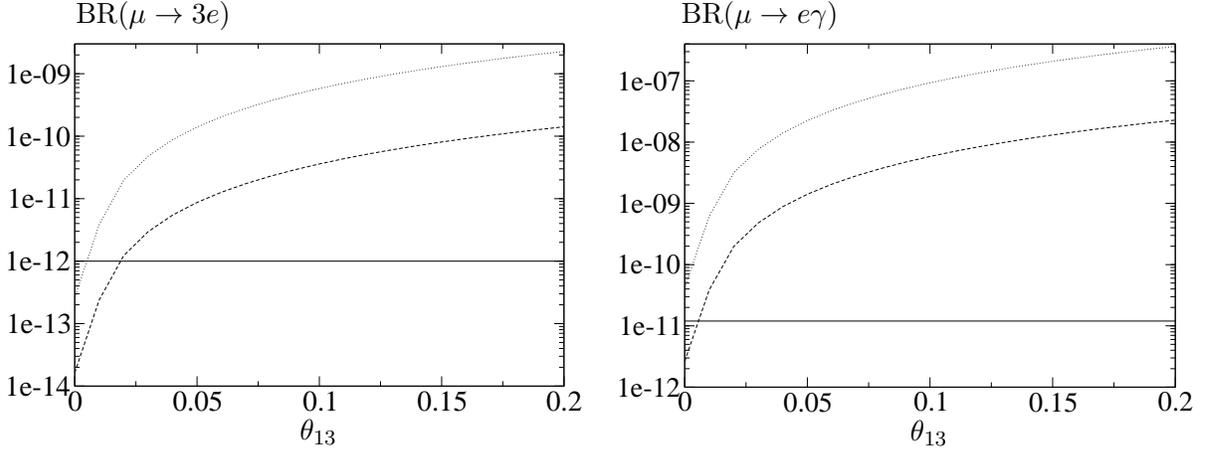

\begin{center}
\psfig{figure=fig5_mu3e.epsi,height=3.0in,angle=-90}
\hspace{0.25cm}
\psfig{figure=fig5_muegamma.epsi,height=3.0in,angle=-90}
\end{center}
\caption{Dependence of LFV $\mu$ decays with $\theta_{13}$ in radians with hierarchical heavy neutrinos and $R = 1$, for 
$(m_{N_1},m_{N_2},m_{N_3})=(10^8, 2 \times 10^8, 10^{14})$ GeV, 
$\tan \beta = 50$, $A_0 = 0$ and 
$\mbox{sign}(\mu) > 0$. The upper lines are for $M_0=250 $ GeV, $M_{1/2}=150 $ GeV 
and the lower lines
are for $M_0=400 $ GeV, $M_{1/2}=300 $ GeV. 
The horizontal lines are the experimental bounds.
\label{fig:5}}
\end{figure}

For the more interesting hierarchical case we show in Figs.~\ref{fig:1}-\ref{fig:5} the numerical results for the branching ratios of the LFV $\tau$ and $\mu$ decays for both types, $l_j \to 3 l_i$ and the correlated radiative decays $l_j \to l_i \gamma$. All the LFV $\tau$ and $\mu$ decay rates are mainly sensitive to $\tan{\beta}$, the heaviest neutrino mass $m_{N_3}$, which we have set in these figures to $m_{N_3} = 10^{14}$ GeV, and the complex angles in the $R$ matrix $\theta_1$ and $\theta_2$, which have been taken in the range $3 < \tan{\beta} < 50$, $0 < |\theta_i|< 3$ and $0 < \arg(\theta_i) < \pi/4$.
In Fig.~\ref{fig:1} we show separately the various contributions to 
BR$(\tau^- \to \mu^- \mu^- \mu^+)$ as a function of $|\theta_2|$.
The dominant one is the photon-penguin contribution (which is
undistinguisible from the total in this figure) and the 
others (Z-penguin, H-penguin and boxes) are several orders of magnitude smaller. All the rates for $\tau^- \to \mu^- \mu^- \mu^+$ in this plot are within the allowed range by the experimental bound, which is placed just at the
upper line of the rectangle. In contrast, we have obtained that, for the mSUGRA and seesaw parameters of Fig.~\ref{fig:1}, the predicted BR$(\tau \to \mu \gamma)$ is above the experimental bound for the explored range of $0 < |\theta_2| < 3$. 
\begin{figure}[t!]
\vspace{-1.5cm}
\begin{center}
\psfig{figure=fig2_tau3mu.epsi,height=2.5in,angle=-90}
\psfig{figure=fig2_taumugamma.epsi,height=2.5in,angle=-90}\\
\vspace{0.5cm}
\psfig{figure=fig2_tau3e.epsi,height=2.5in,angle=-90}
\psfig{figure=fig2_tauegamma.epsi,height=2.5in,angle=-90}\\
\vspace{0.5cm}
\psfig{figure=fig2_mu3e.epsi,height=2.5in,angle=-90}
\psfig{figure=fig2_muegamma.epsi,height=2.5in,angle=-90}
\end{center}
\caption{Dependence of LVF $\tau$ and $\mu$ decays with 
$\vert \theta_2 \vert$ with hierarchical heavy neutrinos 
and complex R, for $\arg(\theta_2) = 0, \pi/10, \pi/8, \pi/6, \pi/4$ in radians 
(lower to upper lines), $(m_{N_1},m_{N_2},m_{N_3})=(10^8, 2 \times 10^8, 10^{14})$ GeV, 
$\theta_1=\theta_3=0$, $\tan \beta = 50$, $M_0=400 $ GeV, $M_{1/2}=300 $ GeV, 
$A_0 = 0$ and $\mbox{sign}(\mu) > 0$. The horizontal lines are the upper experimental bounds.
\label{fig:2}}
\end{figure}
In Fig.\ref{fig:2} we show the predictions of BR$(l_j^- \to l_i^- l_i^- l_i^+)$ and BR$(l_j \to l_i \gamma)$ as functions of $\vert \theta_2 \vert$, for all the channels and for different values of $\arg(\theta_2)$. It is clear that all the branching ratios have a soft behaviour with $\vert \theta_2 \vert$ except for the case of real $\theta_2$ where appears a narrow dip in each plot. In this Fig.~\ref{fig:2} we see that all the rates obtained are below their experimental upper bounds, except for the processes $\tau \to \mu \gamma$ and $\mu \to e \gamma$, where the predicted rates for complex $\theta_2$ with large $\vert \theta_2 \vert$ are clearly above the allowed region. The most restrictive channel in this case is $\tau \to \mu \gamma$ where compatibility with data occurs just for real $\theta_2$ and for complex $\theta_2$ but with $\vert \theta_2 \vert$ values near the region of the narrow dip. We also see that the rates for $BR(\mu \to 3 e)$ enter in conflict with experiment at the upper corner of large $\vert \theta_2 \vert$ and large $\arg(\theta_2) = \pi/4$.
\begin{figure}
\vspace{-1.5cm}
\begin{center}
\psfig{figure=fig3_tau3mu.epsi,height=2.5in,angle=-90}
\psfig{figure=fig3_taumugamma.epsi,height=2.5in,angle=-90}\\
\vspace{0.5cm}
\psfig{figure=fig3_tau3e.epsi,height=2.5in,angle=-90}
\psfig{figure=fig3_tauegamma.epsi,height=2.5in,angle=-90}\\
\vspace{0.5cm}
\psfig{figure=fig3_mu3e.epsi,height=2.5in,angle=-90}
\psfig{figure=fig3_muegamma.epsi,height=2.5in,angle=-90}
\end{center}
\caption{Dependence of LFV $\tau$ and $\mu$ decays with 
$\vert \theta_1 \vert$ with hierarchical heavy neutrinos 
and complex R, for $\arg(\theta_1) = 0, \pi/10, \pi/8, \pi/6, \pi/4$ in radians
(lower to upper lines), $(m_{N_1},m_{N_2},m_{N_3})=(10^8, 
2 \times 10^8, 10^{14})$ GeV, $\theta_2=\theta_3=0$, $\tan \beta = 50$, 
$M_0=400 $ GeV, $M_{1/2}=300 $ GeV, $A_0 = 0$ and $\mbox{sign}(\mu) > 0$. The horizontal lines are 
the upper experimental bounds.
\label{fig:3}}
\end{figure}
Even more interesting are the predictions for 
$BR(l_j^- \to l_i^- l_i^- l_i^+)$ and $BR(l_j \to l_i \gamma)$ as functions of
 $\vert \theta_1 \vert$, due
to the large values of the relevant entries of the $Y_{\nu}$ coupling
matrix. Concretely, $|Y_{\nu}^{13}|$ can be as large as $\sim 0.2$ for $|\theta_1|
\sim 2.5$ and $\arg{(\theta_1)} = \pi/4$, and $|Y_{\nu}^{23}|$ and
$|Y_{\nu}^{33}|$ are in the range $0.1 - 1$ for all studied complex $\theta_1$ values. The results for $BR(l_j^- \to l_i^- l_i^- l_i^+)$ and $BR(l_j \to l_i \gamma)$ as  functions of $\vert \theta_1 \vert$, for different values of $\arg{(\theta_1)}$, are illustrated in Fig.~\ref{fig:3}. We see clearly that the restrictions are more severe in this case than in the previous one. In fact, all the rates cross the horizontal lines of the experimental bounds except for $BR(\tau^- \to \mu^- \mu^- \mu^+)$ and $BR(\tau^- \to e^- e^- e^+)$. The most restrictive channel is now the $\mu \to e \gamma$ decay. More specifically, we see that all the points in the plot of $BR(\mu \to e \gamma)$, except for the particular values $\theta_1= 0$ and real $\theta_1$ at the dip, are excluded by the experimental upper bound. Also the predictions for $BR(\mu \to 3e)$ are mostly excluded, except again for the region close to zero and the dip. We have also explored the dependence with the complex $\theta_3$ angle, and it turns out that the predictions for all rates are nearly constant with this
angle.
We have also tried another input values for the heavy neutrino masses.
The results for $BR(\tau \to 3 \mu)$ are shown in Fig.~\ref{fig:4}. Here we
compare the predictions for the three following set of values,
$(m_{N_1},m_{N_2},m_{N_3})=(10^8, 2 \times 10^8, 10^{14})$ GeV,
$(10^{10}, 2 \times 10^{10}, 10^{14})$ GeV and 
$(10^8, 2 \times 10^8, 10^{12})$ GeV. We conclude, that the relevant mass is the
heaviest one, $m_{N_3}$, and the scaling with this mass is approximately as 
the scaling with
the common mass $m_N$ in the  degenerate case, namely, $(m_{N3} \log{m_{N3}})^2$. Because of this, the rates for 
the two first sets are nearly undistinguisable, and the rates for the third set
are about four orders of magnitude below.
Finally, we consider the very interesting case where the 
angle $\theta_{13}$ of the $U_{MNS}$ is non vanishing. It is known that the
present neutrino
data still allows for small values of this angle, $\theta_{13}<10^o$. 
The dependence of 
$BR(\mu^- \to e^- e^- e^+)$ and $BR(\mu \to e \gamma)$ with this 
$\theta_{13}$ is shown in Fig.~\ref{fig:5} where we explore values in the
$0 < \theta_{13} < 10^o$ range. We choose these two channels
because they are very sensitive to this angle and because they have the most restrictive experimental bounds. For this study we 
assume the simplest choice of $R = 1$, and set the other parameters 
to the following values: 
$\tan \beta = 50$, $A_0 = 0$, $\mbox{sign}(\mu) > 0$, and 
$(m_{N_1},m_{N_2},m_{N_3})=(10^8, 2 \times 10^8, 10^{14})$ GeV. The upper lines
are for $M_0=250 $ GeV, $M_{1/2}=150 $ GeV and the lower ones for
$M_0=400 $ GeV, $M_{1/2}=300 $ GeV.
We conclude that, for this choice of parameters, values of $\theta_{13}$ 
in the upper region of the interval $0 < \theta_{13} < 10^o$ are clearly disfavored by the data on LFV $\mu$ decays.
It is a quite stricking result.

In summary, we obtain in the hierachical case much 
larger rates than in the degenerate one, 
and one must pay attention to these values, 
because the rates in several channels do get in conflict with the 
experimental bounds. More specifically, the choice of a complex $R$ matrix 
 with large modules and/or large arguments of $\theta_1$ and/or $\theta_2$ 
and a light SUSY spectrum is very constrained by data. 
We also confirm that the experimental upper bounds of the 
processes 
$l_j \to l_i \gamma$ are more restrictive than the 
$l_j^- \to l_i^- l_i^- l_i^+$ ones but all together will allow to 
extract large excluded regions of the mSUGRA and seesaw parameter space. 
A more precise conclusion on the excluded regions of this parameter space 
deserves a more devoted study.
\section*{Acknowledgments}
E. Arganda wishes to thank J.M. Frere for the invitation to the conference and all the organizers for a very fruitful, interesting and enjoyable meeting.

\end{document}